\documentclass[sigconf]{acmart}
\AtBeginDocument{%
  }

\setcopyright{acmlicensed}
\copyrightyear{2018}
\acmYear{2018}
\acmDOI{XXXXXXX.XXXXXXX}
\acmConference[ASE 2026]{The 41st IEEE/ACM International Conference on Automated Software Engineering}{October 12--16, 2026}{Munich, Germany}
\acmISBN{978-1-4503-XXXX-X/2018/06}

\usepackage{amsmath}    
\usepackage{amssymb}    
\usepackage{algorithm}  
\usepackage{algpseudocode} 
\usepackage{multirow}
\usepackage{makecell}
\usepackage{tabularx}
\usepackage{float}    
\usepackage{etoolbox}
\setcopyright{none}
\renewcommand\footnotetextcopyrightpermission[1]{}

\usepackage{tikz}
\usepackage{tcolorbox}
\usepackage[table]{xcolor}             

\definecolor{deepBlue}{RGB}{0, 82, 204}      
\definecolor{deepGreen}{RGB}{0, 120, 60}     
\definecolor{deepRed}{RGB}{200, 30, 30}      
\definecolor{deepGray}{RGB}{100, 110, 120}   
\definecolor{deepTeal}{RGB}{0, 128, 128}     
\definecolor{widgetBg}{RGB}{248, 249, 250}   
\definecolor{metaColor}{RGB}{80, 80, 80}     
\definecolor{lightgray}{gray}{0.95}          
\definecolor{headergray}{gray}{0.92}         
\usepackage{listings}
\usepackage{xcolor}

\definecolor{jsonKey}{RGB}{127, 0, 85}    
\definecolor{jsonString}{RGB}{42, 0, 255} 
\definecolor{jsonPunct}{RGB}{102, 102, 102}
\definecolor{jsonComment}{RGB}{60, 179, 113}

\lstdefinelanguage{JSON}{
    basicstyle=\ttfamily\small,          
    numbers=left,                        
    numberstyle=\tiny\color{gray},       
    frame=single,                        
    breaklines=true,                     
    showstringspaces=false,              
    literate=
     *{0}{{{\color{jsonNumber}0}}}{1}    
      {1}{{{\color{jsonNumber}1}}}{1}
      {2}{{{\color{jsonNumber}2}}}{1}
      {3}{{{\color{jsonNumber}3}}}{1}
      {4}{{{\color{jsonNumber}4}}}{1}
      {5}{{{\color{jsonNumber}5}}}{1}
      {6}{{{\color{jsonNumber}6}}}{1}
      {7}{{{\color{jsonNumber}7}}}{1}
      {8}{{{\color{jsonNumber}8}}}{1}
      {9}{{{\color{jsonNumber}9}}}{1}
      {:}{{{\color{jsonPunct}{:}}}}{1}    
      {,}{{{\color{jsonPunct}{,}}}}{1}    
      {[}{{{\color{jsonPunct}{[}}}}{1}    
      {]}{{{\color{jsonPunct}{]}}}}{1},   
    morestring=[b]",                     
    stringstyle=\color{jsonString},      
    morekeywords={true,false,null},      
    keywordstyle=\color{jsonKey},        
    comment=[l]{\#},                     
    commentstyle=\color{jsonComment},
    identifierstyle=\color{jsonKey}      
}
\usepackage{tikz}
\begin{document}

\title{An Automated Framework for Input Alphabet Construction in Stateful Protocol Implementation Learning}

\author{JiongHan Wang}
\email{wangjh1@mail.ustc.edu.cn}
\affiliation{%
  \institution{University of Science and Technology of China}
  \city{He Fei}
  \country{China}
}
\author{WenChao Huang}
\email{huangwc@mail.ustc.edu.cn}
\affiliation{%
  \institution{University of Science and Technology of China}
  \city{He Fei}
  \country{China}
}

\renewcommand{\shortauthors}{Trovato et al.}

\begin{abstract}


As a prevalent analytical technique for stateful protocol implementations, state machine learning suffers from a core bottleneck stemming from handcrafted input alphabets. Manual alphabet definition inherently limits the completeness of input exploration, making it difficult to capture anomalous non-conformant messages and consequently missing latent semantic defects.

In this paper, we target automatic input alphabet generation to break the above limitation for state machine learning. We adopt large language models to parse protocol message layouts and produce candidate input symbols following structured mutation rules, which automatically covers valid and invalid message spaces and eliminates reliance on manual protocol expertise.

Considering the rising overhead brought by continuously growing alphabets, we introduce a mini-batch incremental learning strategy to reuse existing learned automata when incorporating new alphabet entries. Comprehensive experiments on practical protocol stacks indicate our approach can reproduce existing security vulnerabilities and identify novel semantic bugs. A subset of these newly discovered issues has been confirmed and patched by developers, proving the practicability and effectiveness of our proposed method.
\end{abstract}


\maketitle

\section{Introduction}

Stateful protocol implementations are an important component of computer systems, and the correctness of their execution semantics is critical for the stable operation of real-world systems. For many important protocols, the behavior of protocol deployments is specified by protocol specification documents, and their semantic correctness means that the externally observable behavior at runtime is fully consistent with the request-response semantics defined by the protocol specification (RFC documents).
If a protocol implementation contains semantic bugs, it can severely impact their functionality and security. For example, the TLS protocol and its deployment guarantee information security in most network communications~\cite{maehren2025towards}. Vulnerabilities such as EarlyCCS~\cite{beurdouche2015messy} found in TLS deployments can severely compromise its security.

Semantic bugs in stateful protocol implementations continue to emerge as protocols and their deployments grow in scale~\cite{natella2021profuzzbench}. Existing approaches for detecting such bugs include formal verification, model checking, fuzzing, and state machine learning. Formal verification can validate protocol specifications but typically only scales to relatively simple protocols and requires substantial expert effort~\cite{arquint2023generic,arquint2023sound}, while model checking similarly involves significant manual intervention~\cite{ferreira2021prognosis}. Fuzzing has been widely applied to protocol testing (e.g., AFLNet~\cite{pham2020aflnet}), but most approaches primarily target memory corruption vulnerabilities rather than semantic bugs, and recent tools such as DYFuzzing~\cite{Ammann2024DYFuzzing} still require complex instrumentation and expert-defined properties. By comparison, state machine learning provides a relatively convenient way to analyze protocol behaviors and has been applied to many critical protocols~\cite{maehren2025towards, deRuiter2015Learning, fiterau2020analysis}.

Despite its advantages in ease of deployment and its natural suitability for semantic bug analysis, state machine learning suffers from a major limitation: existing algorithms operate only on a manually defined input alphabet. The input alphabet is a collection of input symbols, where each input symbol serves as an abstraction of concrete messages transmitted to the system under test. As a result, if the messages corresponding to the defined alphabet cannot cover those required to trigger vulnerabilities, any learning strategy will inevitably fail to expose such bugs. From a technical perspective, this gap introduces two key challenges.

\textbf{Challenge 1: Automatically obtaining useful input symbols.}
Existing state machine learning approaches rely heavily on expert knowledge to manually define input symbols~\cite{tranvan2024mealy,Baumer2025FindingSSH,maehren2025towards,Rasoamanana2022Systematic}. This process requires prior knowledge of both the protocol specification and its implementation. Moreover, for complex protocols designing specialized symbols and ensuring that they can be translated into valid, and executable messages is itself a labor-intensive task.

\textbf{Challenge 2: Balancing the trade-off between alphabet size and learning overhead.}
When facing an extended input symbol set with dozens or even hundreds of elements, constructing a compact and effective alphabet from it remains a non-trivial challenge. On the one hand, the effectiveness of state machine learning depends on the alphabet: semantic bugs can only be detected if the alphabet contains symbols capable of triggering them, which encourages using a larger alphabet. On the other hand, state machine learning is computationally expensive~\cite{angluin1987learning}, and excessive alphabet expansion can significantly increase the learning cost or even prevent the learning process from terminating.

To address the challenges of existing state machine learning methods, this paper designs and implements a novel framework that can automatically construct an extended input symbol set and extract alphabets capable of supporting efficient protocol semantic vulnerability detection.
Specifically, it is developed based on two key observations: First, recent advances in pre-trained large language models (LLMs) have shown that LLMs can understand the structure of protocol messages and convert them into a unified, machine-parsable standard format~\cite{meng2024large}. We leverage this capability to generate protocol message configurations for constructing basic symbols. Combined with a structured mutation strategy, it can automatically generate richer and more diverse candidate input symbols for state machine learning. Second, we observe that protocol state machines usually exhibit stable operating characteristics: except for a small number of critical inputs, most input symbols only perform state-local operations and do not trigger cross-state transitions. Taking advantage of this property, we propose and implement Mini-Batch Learning, a method that extracts a set of extended alphabets from the expanded input symbol set, combined with a state machine learning algorithm equipped with a built-in basic state machine.

Experimental results demonstrate that our method can automatically construct extended input sets and derive alphabets for various protocols, and further complete state machine learning on these alphabets to detect semantic bugs. In comparison, the original unextended input symbol sets fail to cover the message types required to trigger multiple real-world semantic bugs. Moreover, naive state machine learning directly adopting the extended input set as the overall alphabet cannot terminate within a reasonable time budget when evaluated on several protocol implementations. As a result, Our approach successfully reproduces two classic high-severity historical semantic vulnerabilities and discovers ten novel semantic bugs in total. Among these newly identified defects, three have been confirmed and patched by official developers, and one has been assigned an CVE identifier.

In summary, this paper makes the following contributions:

\textbf{1. Mutation-driven state machine learning for semantic bug detection:}
We propose a mutation-driven methodology that integrates message mutation with state machine learning. By leveraging large language models to  protocol message configurations, our approach enables protocol-agnostic and systematic exploration of protocol behaviors.

\textbf{2. Efficient learning over the set of extended alphabets:}
We propose and deploy a mini-batch learning strategy. By partitioning the input symbol set into a collection of extended alphabets and integrating it with a learning algorithm equipped with a baseline state machine, we reduce the number of membership queries required during learning, thereby ensuring practical termination even with large input symbol sets.

\textbf{3. Practical evaluation and bug discovery:}
We evaluate our approach on multiple implementations of widely deployed stateful protocols. Our method discovers previously unknown semantic bugs, three of which have been fixed by developers and one assigned a CVE identifier.

\section{Background}
\textbf{State Machine Learning:} 
Active automata learning has been widely used to infer behavioral models of protocol implementations. This line of work is rooted in Angluin’s $L^*$ algorithm~\cite{angluin1987learning}, which learns automata through membership and equivalence queries. In this work, we adopt state machine learning techniques to synthesize the input–output behavior of a protocol implementation into a Mealy machine. A Mealy machine models the protocol as a tuple 
$SM=(Q, q_0, \Sigma, O, \delta, \lambda)$, where $Q$ denotes the set of states, $q_0$ is the initial state, $\Sigma$ is the input alphabet, $O$ is the output alphabet, and $\delta$ and $\lambda$ represent the transition and output functions, respectively~\cite{angluin1987learning}. An alphabet is a set of symbols, where each symbol represents an abstraction of a specific sent or received message.

\textbf{From Learned State Machines to Semantic Bug Detection:}
The learned state machine provides an abstract representation of the protocol implementation's behavior and can reveal deviations from the intended protocol logic. However, identifying implementation flaws from the learned model remains a challenging task. While formal verification techniques can in principle analyze properties of finite-state models, exhaustive verification often suffers from the state explosion problem, making fully automated analysis computationally expensive in practice.

Consequently, prior work on protocol state machine learning has typically relied on manual inspection of the inferred models to identify unexpected transitions or states~\cite{Rasoamanana2022Systematic,StateInspector2022,deRuiter2015Protocol}. Other approaches design protocol-specific analysis techniques that detect particular classes of implementation flaws based on domain knowledge~\cite{ferreira2021prognosis,tranvan2024mealy}. While effective in certain scenarios, these approaches often require significant expert effort and are difficult to generalize across different protocols.

\section{Motivating Example}
\label{sec:Motivation Example}


\begin{figure}[t]
  \centering
  \includegraphics[width=0.8\linewidth]{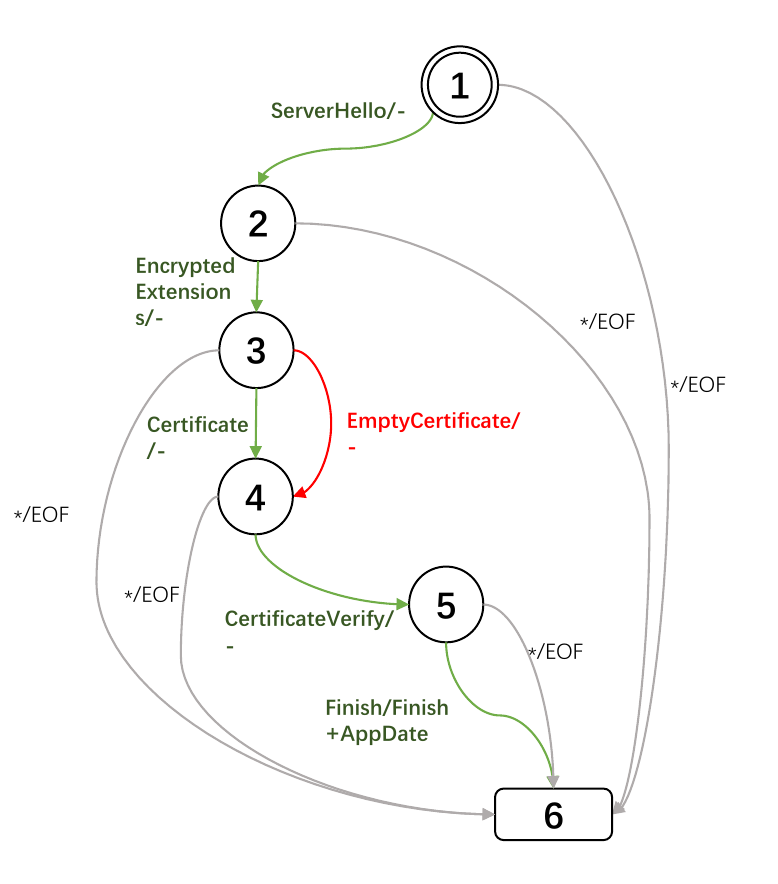}
  \caption{CVE-2022-25638, Vulnerable State Machine}
  \label{fig:cve2022}
\end{figure}

We adopt CVE-2022-25638~\cite{Rasoamanana2022Systematic}, a critical semantic bug in the TLS 1.3 stack of WolfSSL, as our motivating example throughout this work. As shown in Figure \ref{fig:cve2022}, an attacker can first send an empty certificate message when a certificate and verification message are expected, and then follow it with an invalid certificate verification message (containing an unknown signature algorithm and arbitrary payload), thereby bypassing server authentication. The core condition for triggering this vulnerability is to introduce an Empty Certificate message (EmptyCert) into the message flow together with a carefully crafted invalid CertificateVerify message (CV\_invalid).

However, acquiring the two key messages to trigger the vulnerability is non-trivial. The EmptyCert message does not exist in the standard message specification of TLS 1.3 servers~\cite{rfc8446}. Injecting such a message into a state machine learning or testing workflow therefore requires additional expert knowledge. Constructing the invalid CertificateVerify message is even more challenging for general testers. This message contains eight subfields in total. To trigger the vulnerability, the signature algorithm field must contain a value that violates the specification, while all other fields must remain valid. Meanwhile, the constraints of both the record layer and the handshake layer headers must still be satisfied. Such a process demands deep protocol expertise and careful manual crafting. Consequently, vulnerability discovery methods that rely heavily on expert knowledge are difficult to generalize across different protocols or even across different implementations of the same protocol. This observation motivates us to explore methods that can automatically construct non-compliant messages for state machine learning and testing.


These challenges motivate the core research objective of this work: automatically constructing an alphabet that covers non-standard and non-compliant protocol messages. By leveraging the knowledge and automated reasoning capabilities of Large Language Models, our system aims to operate across different protocols and implementations.



\begin{figure}[t]
  \centering
  \includegraphics[width=\linewidth]{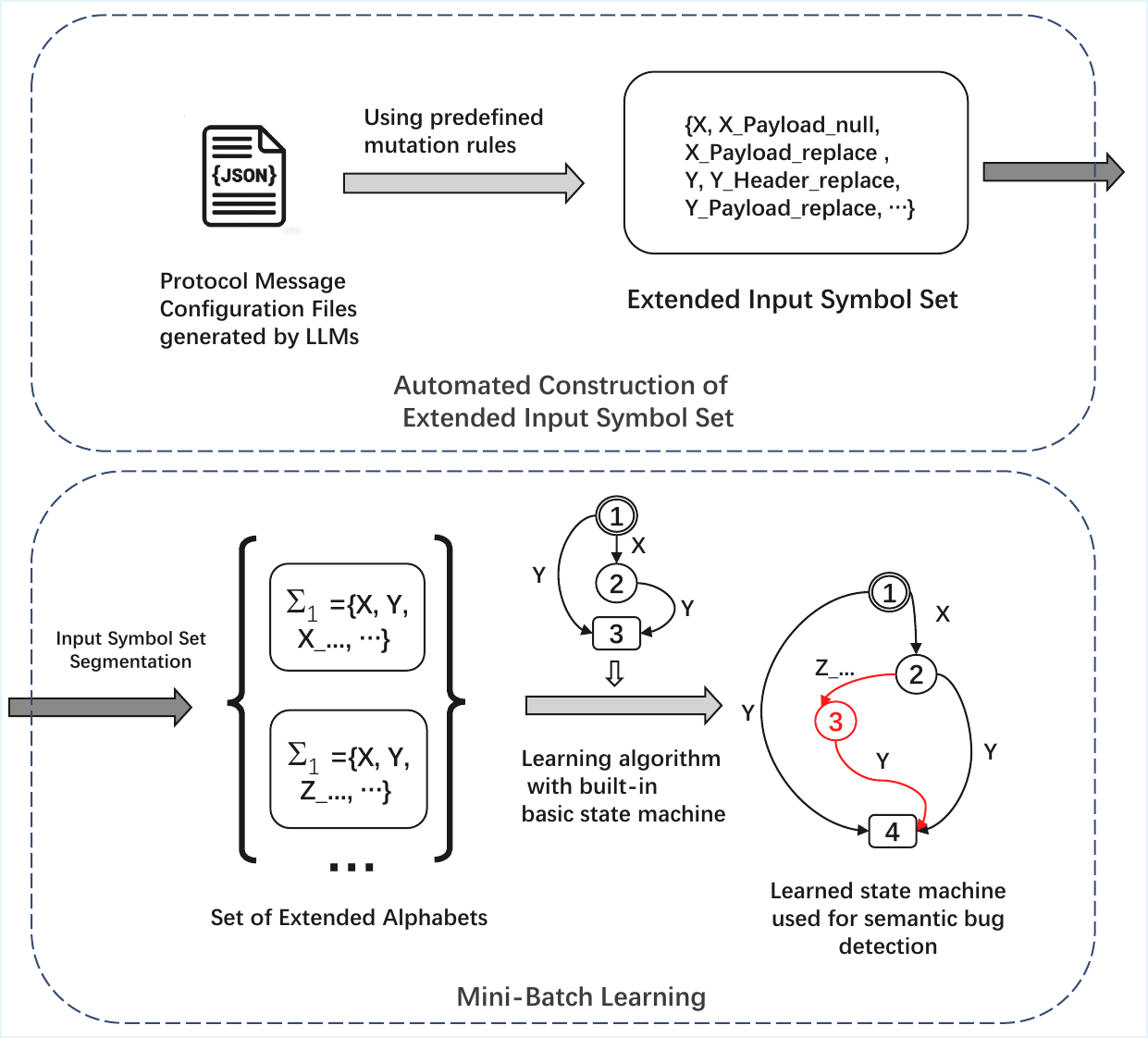}
  \caption{Framework Overview}
  \label{fig:overview}
\end{figure}
\section{Methodology}
\subsection{Overview}

Our method aims to automatically construct input alphabets for state machine learning of the system under test, thereby achieving efficient semantic vulnerability detection. As shown in Figure \ref{fig:overview}, this method involves two key steps: obtaining the extended input symbol set and performing mini-batch learning. The first step involves acquiring the extended input symbol set, which requires using a large language model to extract protocol message format configurations. Based on mutation rules, new mutated input symbols are then generated to form the extended input symbol set. The second step is to perform mini-batch learning on the extended input symbol set to obtain a state machine that describes the system's input-output behavior. This process requires organizing the extended input symbol set into a collection of extended alphabets, which is then used with the newly designed state machine learning method.

\subsection{Automated Construction of 
the Extended Input Symbol Set}

We first need to construct abstract symbols and define how they can be translated into concrete messages for transmission. Traditionally, defining a finite input alphabet for state machine learning relies heavily on expert knowledge, where protocol-specific alphabets are manually constructed based on a deep understanding of the target protocol~\cite{maehren2025towards}. This process is labor-intensive and lacks generalizability across different protocols. To address this limitation, we propose an LLM-augmented approach inspired by fuzzing strategies, which constructs the input space through finite mutations of valid messages derived from protocol specifications.

Specifically, we decompose the alphabet construction problem into three steps: (1) extracting structural information of protocol messages; (2) generating an extended set of symbols by mutating messages based on their structural properties; and (3) constructing a message suite that maps abstract symbols to the concrete messages sent during protocol interactions.


\subsubsection{LLM-assisted protocol message configuration acquisition:} We leverage the capabilities of LLMs in aggregating and parsing protocol information to generate machine-readable configuration files. First, we abstract protocol messages into a Header-Payload structure to establish a foundation for subsequent instruction description generation. The header generally represents the command opcode; for instance, in the FTP command "USER ubuntu", "USER" serves as the header and "ubuntu" as the payload. Additionally, we introduce a comment field to encapsulate supplementary information. Specifically, we provide two types of metadata: Enumeration values for the payload (e.g., the TYPE command in FTP may accept values such as A, I, E, or L) and Relationships with data connections. In protocols like FTP and RTSP, where communication involves both control and data channels, certain commands require explicit specification of their interaction with the data connection. Based on these abstract data types, we propose an LLM-augmented solution that synthesizes protocol specifications to construct standardized JSON descriptions. These JSON files encapsulate message types, field constraints, valid value ranges, and mappings to abstract data types. The prompt design employs a few-shot learning approach, utilizing curated examples to guide the LLM in generating structurally consistent JSON content, thereby minimizing human intervention during the message generation process. 

Subsequent experiments demonstrate that large language models are fully capable of handling the task of constructing message configurations. The message configurations generated by the large language model comply with established grammatical specifications. These configurations can be parsed by pre-designed programs and have been verified manually to conform to protocol specifications. The program-readable configuration files they provide for different protocols lay the foundation for the cross-protocol execution of our method.

\nopagebreak
\begin{lstlisting}[language=JSON, label=code:ftp, float=tbh, keepspaces,caption={JSON Definition of FTP TYPE Command}]
{
    "command_code": "TYPE",
    "description": {
        "header": "TYPE",
        "payload": "A ",
        "comment": "enumerable valid values: [\"I\", \"E\", \"L\"]; connection status: no connection required"
    }
}
\end{lstlisting}

For example, for the TYPE message in the FTP protocol as shown above, our approach can generate the following entry in the message structure configuration file which covers all information of the message, from the command code and possible parameters to state requirements for the data connection. The subsequent message sending program can automatically construct and send the TYPE message and its mutated response messages based on this configuration.

\subsubsection{Fault Classification-Guided Mutation:}\label{sec:Fault Classification-Guided Mutation} Based on different fault types, we design three mutation strategies to extend specification-compliant messages, thereby constructing the extended input symbol set. We use the abstract data types (Header-Payload structures) and their corresponding JSON instruction files generated in the previous step as the basis for mutation operations to construct customized mutation strategies for subsequent state machine learning. We propose three mechanically implementable mutation strategies as follows:

\begin{itemize}
\item {\texttt{1, Payload Nullification}}: delete the payload portion of the message; this mutation rule is not applicable to messages that originally have an empty payload.
\item {\texttt{2, Payload Error Content Replacement}}: replace payload fields with error content or non-message-based configurations.
\item {\texttt{3, Header Error Content Replacement}}: replace header fields with error content.
\end{itemize}

It should be noted that for mutation methods 3 and 4, there is actually considerable room for manipulation. For example, we can delete part of the compliant message fields, add new content to them, duplicate the entire message, or directly generate a random value for substitution. All four of these mutation modes are supported in our code, yet they are categorized as a single type during classification. For instance, with the support of the aforementioned mutation methods, all types of TYPE commands we can obtain include: \{\text{TYPE A},~\text{TYPE},~\text{TYPE I},~\text{TYPE E},~\text{TYPE AA},~\text{TYPE df},~\text{TYE A},\\~\text{TYPED A},~\text{TYPETYPE A}\}

In this study, we define the input space as comprising the basic messages derived from RFC documents and the mutated messages generated by our defined strategies; the abstract symbols corresponding to these messages constitutethe extended input symbol set:
\begin{definition}
Let the extended input symbol set be \( U \), the original symbols set be \( P \), and the mutated symbols set be \( M \). Then the full symbols set can be expressed as \( U = P \cup M \). Among them, the mutated symbols set \( M \) satisfies: for any \( m \in M \), there exists a unique \( p \in P \) and a mutation operation \( f \) such that \( m = f(p) \).
\end{definition}
Based on this definition, all subsequent symbols are contained in the set $U$, and the distinction between mutated and non-mutated symbols benefits our mini-batch learning.
\subsubsection{Construct the mapping from symbols to transmitted messages:}To enable interaction between the learning algorithm and a real protocol implementation, a test harness is constructed. The test harness translates abstract input symbols used by the learner into concrete protocol messages and converts the responses from the implementation back into abstract outputs. For simple protocols, this process mainly involves mapping abstract messages to concrete message formats. For more complex protocols, additional functionality may be necessary, such as maintaining session states, managing transport connections, or handling cryptographic operations. When mature open-source implementations or libraries are available (e.g., scapy~\cite{secdev_scapy_2026} for TLS), constructing such a test harness can be significantly simplified.

Building upon the client, we develop a general message construction mechanism capable of handling symbols that include mutation operations. For cryptographic protocols such as TLS, we implement a mechanism in which the client first parses the previously generated message structure configuration file to construct standard messages, and then dynamically applies mutation operations during message generation. For text-based protocols such as FTP, although the above approach remains applicable, their simpler message structures allow for a more lightweight solution. Specifically, based on the previously generated message structure configuration file, we construct message-sending configuration files for all symbols, including those with mutation operations. During execution, the client only needs the message-sending configuration file to handle mutated symbols in the same way as regular symbols when sending messages.

\begin{figure}[b!]  
  \raggedright     
  \includegraphics[width=0.48\textwidth]{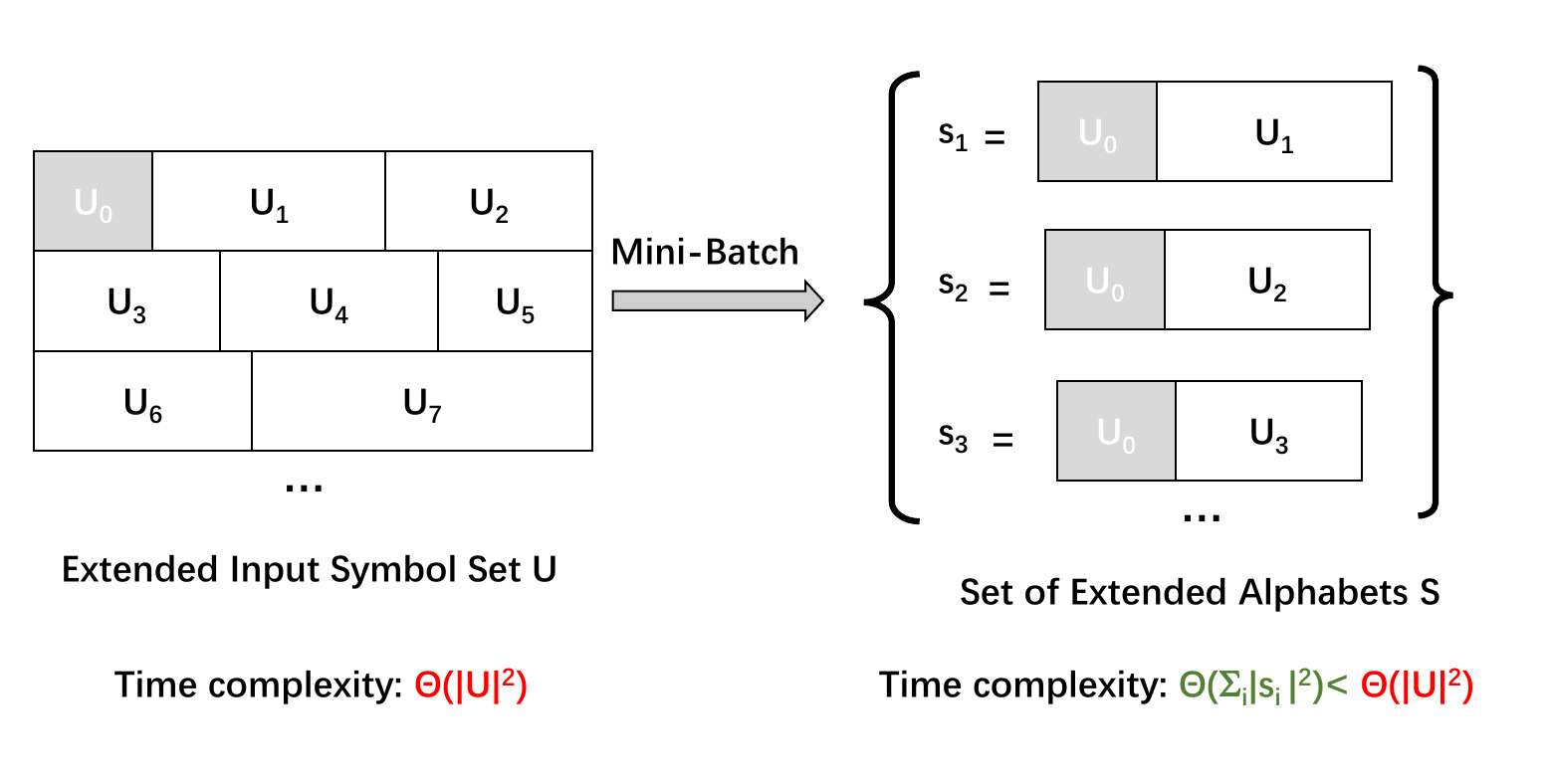}
  \caption{Mini-Batch Learning Strategy}
  \label{fig:Mini-Batch Learning}
\end{figure}
\subsection{Mini-Batch Learning}

We propose a mini-batch learning strategy as shown in Figure \ref{fig:Mini-Batch Learning} to address the issue that the mutation-based approach may produce an excessively large input symbol set that exceeds the practical capability of state machine learning algorithms. Starting from a basic alphabet, we incrementally construct a sequence of extended alphabets. Instead of learning over the entire input symbol set at once, we perform learning on each of these extended alphabets in sequence, thereby reducing the complexity of the learning process. According to the complexity theory of state machine learning, the theoretical lower bound on the number of membership queries for any active learning algorithm is $\Theta(k^2)$, where $k$ denotes the size of the alphabet~\cite{Howar2012Active}.This bound is established with respect to the alphabet size, and does not account for other influencing factors in overall query complexity. With our mini-batch learning approach, this complexity is reduced to $\Theta(\sum k_i^2)$, where $\sum k_i \approx k$. This formulation provides an advantage when $k$ is large. Furthermore, based on this learning strategy, we design a method that incorporates the basic state machine directly into the learning process, which further reduces the overall time overhead.

\subsubsection{Construction of the Set of Extended Alphabets}
We transform the large input symbol set into a collection of extended alphabets suitable for state machine learning by defining the basic alphabet and designing two methods for addition.

To ensure that each alphabet covers all states of the protocol under conforming conditions, we first need to identify the basic alphabet of the protocol. Based on observations of state machines in stateful protocols, we find that many protocol state machines can be viewed as extensions of a core state machine with additional functionalities layered on top. In practice, a protocol often consists of a core state machine augmented with various auxiliary operations or features.

Taking the FTP protocol as an example, a core state machine can be established using commands such as USER, PASS, PASV, PORT, LIST, and QUIT. Most other commands are executed on top of this core state machine to provide specific functionalities, without increasing the number of states in the underlying machine. The alphabet corresponding to this core state machine is defined as the basic alphabet. Under normal circumstances, we can take the set of all ordinary symbols without mutation as the basic alphabet. From a practical perspective, the basic alphabet can be inferred from protocol specification documents as well as prior testing practices.

Once the basic alphabet is established, we can expand other alphabets from it using two heuristic strategies, each motivated by a different idea, as follows.
\begin{figure}[t]
  \centering
  \includegraphics[width=\linewidth]{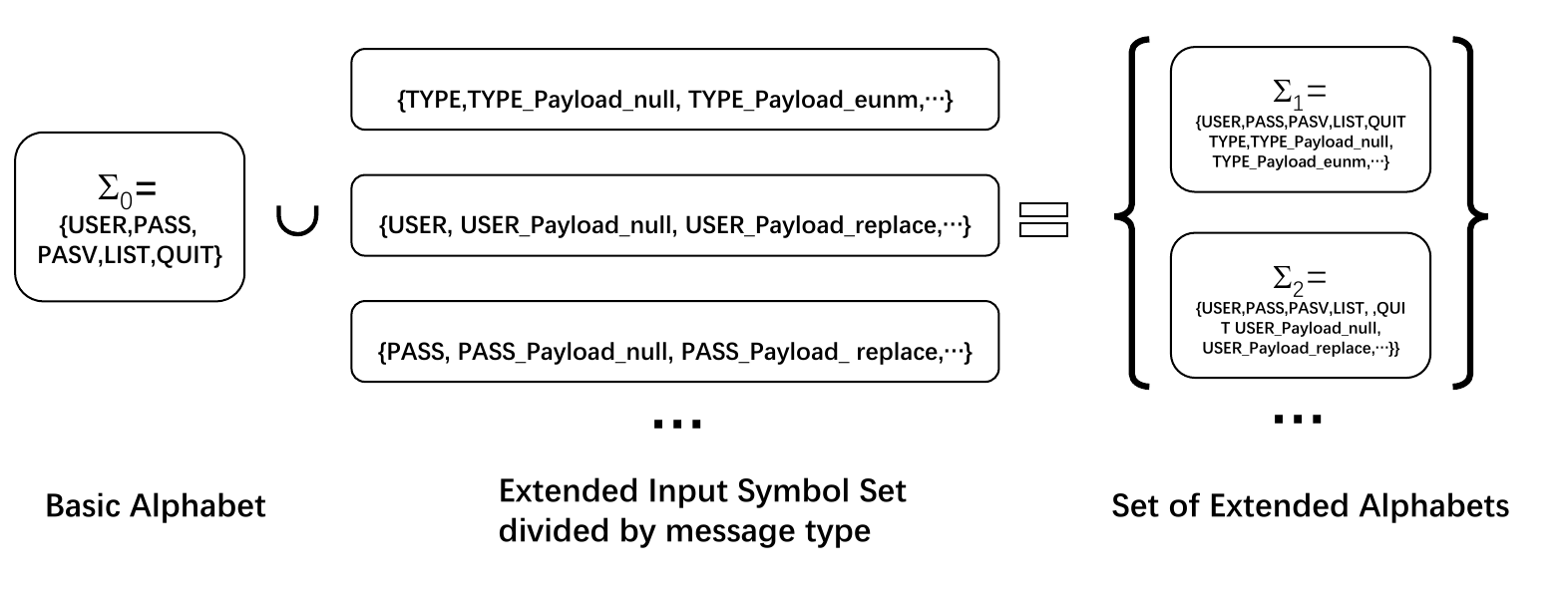}
  \caption{Single-Symbol Mutation Priority}
  \label{fig:alg1}
\end{figure}

\textbf{Single-Symbol Mutation Priority:}
For the target alphabet, select one symbol and add all its possible mutations to the original alphabet, thereby forming a new alphabet. Then, iterate through all the remaining symbols in the alphabet that have not yet been mutated and repeat the process. Implementation details are provided in Algorithm 1 in the appendix.

Taking the FTP protocol as an example, as shown in Figure \ref{fig:alg1}, we partition the set of candidate symbols into multiple subsets according to message types on the basis of our selected basic alphabet. By adding each subset to the basic alphabet separately, we obtain a series of extended alphabets for testing.

\textbf{Diverse-Symbol Mutation Priority:} 
For the target alphabet, prioritize selecting the symbol with the fewest mutations applied, and randomly choose one of its mutations to add to the original alphabet. During the mutation process, a First-In-First-Out (FIFO) active window mechanism is maintained. If the duration of a state machine learning iteration exceeds a predefined threshold or the window size exceeds the maximum limit, the oldest element in the window is discarded. The detailed execution procedure is provided in Algorithm 2 in the appendix.

As shown in Figure \ref{fig:alg2}, in this scheme, several elements are selected from the  input symbol set and added to a queue each time, while a certain number of elements are removed from the queue. The queue is then treated as a set and combined with the basic alphabet to form the extended alphabet.

\begin{figure}[b!]
  \raggedleft       
  \includegraphics[width=0.48\textwidth]{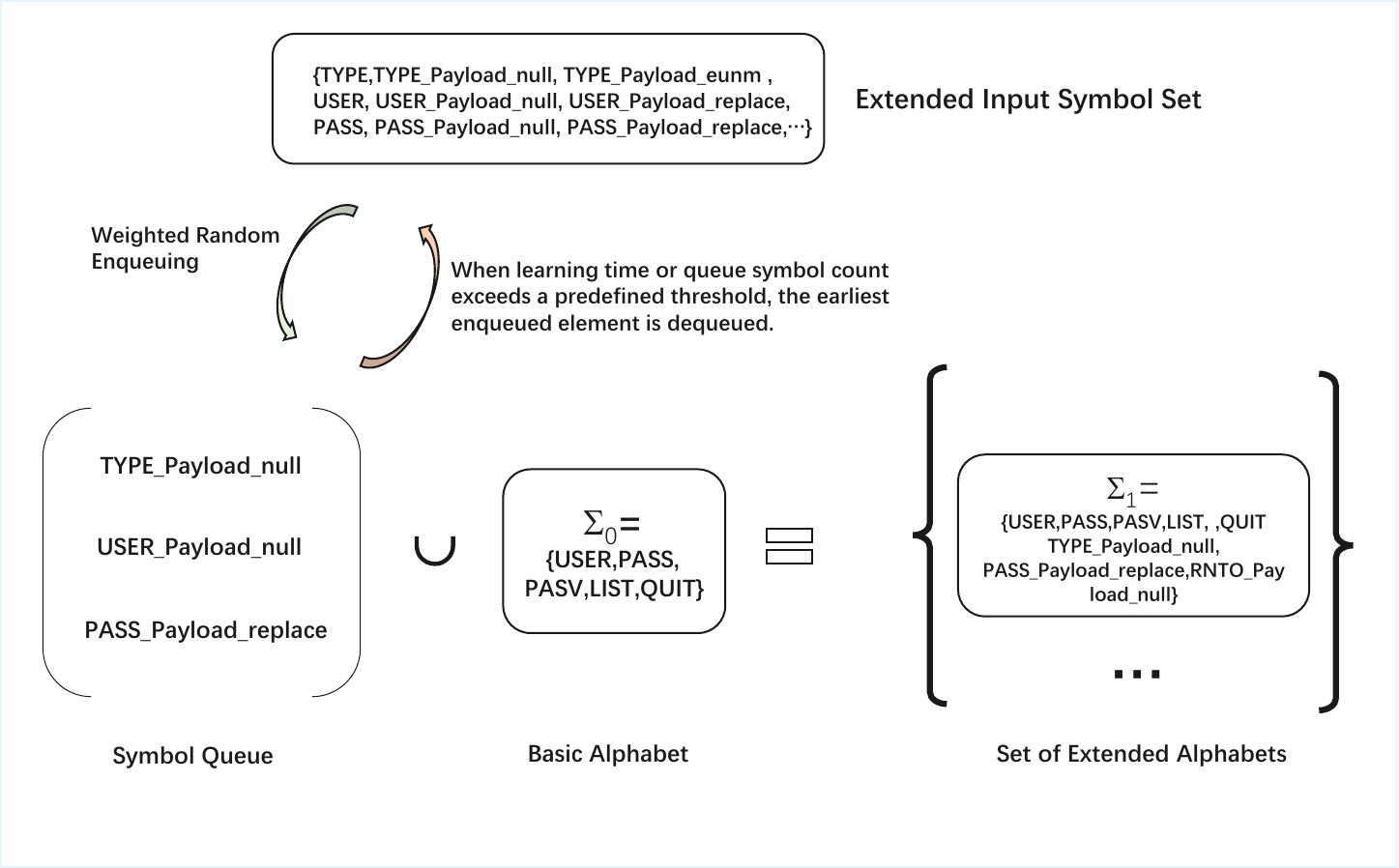}
  \caption{Diverse-Symbol Mutation Priority}
  \label{fig:alg2}
\end{figure}

The two strategies described above correspond to two typical triggering scenarios of semantic bugs: Single-message-type semantic bugs and multi-message-type collaborative semantic bugs. This heuristic prioritization allows us to explore the input space while maintaining the computational tractability of the learning algorithm. Furthermore, this design facilitates subsequent efficiency optimizations and helps us mitigate the time-consuming nature of state machine learning. We define the alphabet set obtained by the above two algorithms as the set of extended alphabets:

\begin{definition}\label{def:set extended alphabet}
Let the set of extended alphabets be \( S \) = \{ $\Sigma_i$,i=1,2,\dots,n \}, and $\Sigma_i$ = $\Sigma_0\cup\Delta_i$
\end{definition}

By defining such a structure, we can clearly observe that the extended alphabets we study exhibit a distinct characteristic: they all contain a subset of the basic alphabet. This observation provides a foundation for our subsequent design of a learning algorithm with a built-in basic state machine.

\subsubsection{Performing State Machine Learning:}

To leverage the fact that each alphabet in the set of extended alphabetss contains the basic alphabet, we propose a method that integrates the known state machine into the learning algorithm, thereby eliminating redundant learning of the basic state machine. Leveraging the structural property defined in Definition~\ref{def:set extended alphabet}, we incorporate the state machine corresponding to $\Sigma_0$ into the learning algorithm, thereby eliminating redundant overhead across multiple learning iterations. The formal algorithm is presented in Algorithm 3 in the appendix.

Intuitively, given an basic alphabet $\Sigma_0$ and the corresponding state machine learning result $SM_0$, when faced with an expanded alphabet $\Sigma_i$ derived from $\Sigma_0$, we exploit the dependencies between observation tables. By reusing the information of the observation table constructed during the learning of $\Sigma_0$, we only need to supplement and verify the table entries corresponding to the expanded portion to complete the state machine learning based on $\Sigma_0$.

The subsequent learning process is consistent with the standard $L^*$ algorithm~\cite{angluin1987learning}. The algorithm maintains an observation table consisting of a set of prefixes, a set of suffixes, and a membership function that records the outputs observed from the SUT. By enforcing the closedness and consistency properties of the observation table, the learner constructs a hypothesis Mealy machine that explains the observed input–output behavior. The learning process alternates between membership queries, which obtain outputs for specific input sequences, and equivalence queries, which check whether the current hypothesis is behaviorally equivalent to the SUT. If the hypothesis is incorrect, a counterexample is returned and used to refine the model. In practical black-box settings where a perfect equivalence oracle is unavailable, counterexamples are typically approximated using testing techniques such as random testing or conformance testing methods. The detailed execution procedure is provided in Algorithm 3 in the appendix.


\begin{table}[htbp]
  \centering
  \caption{Comprehensive Statistics of Tested Subjects}
  \label{tab:protocol_comprehensive_stats}
  \footnotesize
  \begin{tabular*}{\linewidth}{@{\extracolsep{\fill}} l c c c}
    \toprule
    \textbf{Protocol} & \textbf{Subject} & \textbf{\#Stars} & \textbf{Version} \\
    \midrule
    FTP             & LightFTP         & 277              & 5980ea1       \\
                    & ProFTPD          & 581              & 61e621e       \\
                    & Pure-FTPD        & 887              & c21b45f       \\
    \addlinespace[2pt]
    SMTP            & Exim             & 781              & 38903fb       \\
    \addlinespace[2pt]
    RTSP            & Live555          & 851              & 2023.05.10    \\
    \addlinespace[2pt]
    TLS1.3-Server   & wolfSSL          & 2784             & v4.6.0        \\
                    & OpenSSL          & 29847            & 3.0.1         \\
    \addlinespace[2pt]
    TLS1.3-Client   & wolfSSL          & 2784             & v4.6.0        \\
                    & OpenSSL          & 29847            & 3.0.1         \\
    \bottomrule
  \end{tabular*}
\end{table}

\section{Experimental Design}
To thoroughly evaluate the effectiveness of our method, we conducted a series of experiments to address the following research questions:

\textbf{RQ1:} 
With the guidance of large language models, can our method automatically generate and generalize protocol-specific input symbols with high coverage across diverse real-world protocols?

\textbf{RQ2:} 
Can our designed mini-batch learning strategy significantly reduce time overhead and query complexity while maintaining inference accuracy during state machine learning on real-world protocol implementations, compared with conventional learning paradigms?

\textbf{RQ3:}
Can our proposed approach effectively detect protocol implementation inconsistency semantic bugs in state machines when evaluated on real-world protocol implementations?

\subsection{Benchmark}
Table \ref{tab:protocol_comprehensive_stats} presents the subject programs used in our experiments. Our benchmark contains 9 network protocol implementations, covering 5 widely used network protocols, namely RTSP, FTP, SMTP, and the server and client implementations of TLS 1.3. These subject programs cover both cryptographic and non-cryptographic protocols, in the PROFUZZBENCH benchmark~\cite{natella2021profuzzbench} which is a popular benchmark for evaluating stateful protocol fuzzers.
The above protocols cover a variety of application scenarios, including streaming media, messaging, file transfer, and encrypted communication. Their implementations are mature and widely adopted by both enterprises and individual users. Semantic bugs in these projects can have far-reaching impacts.
\subsection{Test Experiment}
All experiments were conducted on a server equipped with an Intel Xeon Platinum 8468V CPU. The machine is configured with 12 logical cores clocked at 2.50 GHz, 16 GB of main memory, and runs on the Windows 10 operating system. In addition, Docker 28.1.1 is deployed to achieve environment isolation and containerized execution of the subject under test.

\subsection{Runtime Configuration}
We standardize key runtime parameters for state machine learning and mutation across all experiments. During the symbol construction phase, we use ChatGPT 5.3 model to generate message format configuration files. Since no existing work has studied the problem of state machine learning over the set of extended alphabets and existing work~\cite{StateInspector2022,Ammann2024DYFuzzing,tranvan2024mealy} cannot automatically switch among the protocols under test, we compare our approach with the basic state machine learning algorithm to demonstrate the effectiveness of our mini-batch learning strategy.

\section{Evaluation}

\begin{table}[htbp]
  \centering
  \caption{Comprehensive Statistics of Tested Protocols}
  \label{tab:protocol_comprehensive_stats}
  \resizebox{\linewidth}{!}{
    \setlength{\tabcolsep}{4pt}
    \begin{tabular}{l c c c c}  
      \toprule
      \textbf{Proto.} & \textbf{Msg\#} & \textbf{Cfg-Spec Msg\#} & \textbf{Sym\#} & \textbf{AlphSz} \\
      \midrule
      FTP             & 34             & 34                   & 207           & 6               \\
      SMTP            & 12             & 12                   & 60            & 12              \\
      RTSP            & 10             & 10                   & 88            & 10              \\
      TLS1.3-Server   & 3              & 3                    & 68            & 3               \\
      TLS1.3-Client   & 6              & 6                    & 113           & 6               \\
      \bottomrule
    \end{tabular}
  }
  \vspace{2pt}
  \scriptsize
  \textit{Notes}: Msg\# = Total number of message types; Cfg-Spec Msg\# = Number of message types defined in configuration files that comply with protocol specifications; Sym\# = Total number of symbols; AlphSz = Basic alphabet size.
\end{table}

\subsection{RQ1: Automated Symbol Construction}
Table \ref{tab:protocol_comprehensive_stats} summarizes the basic characteristics of the LLM-generated configuration files and the constructed input symbol sets. To evaluate the usability of the generated message configuration files, we adopt manual inspection to verify whether the description of each message type in the files conforms to the protocol specifications. As shown in Table \ref{tab:protocol_comprehensive_stats}, all generated configuration files for evaluated protocols comply with the official protocol standards after manual verification, which guarantees the correctness of protocol message processing in the subsequent workflow of our framework.

Based on these configurations, we construct the extended input alphabet by applying the mutation rules described in Section \ref{sec:Fault Classification-Guided Mutation}. Table \ref{tab:protocol_comprehensive_stats} presents the final number of symbols obtained for each protocol.

Regarding the selection of the basic alphabet, all protocols except FTP use the complete set of non-mutated symbols as the basic alphabet. For FTP, however, preliminary experiments and analysis of the specification indicate that fewer than 20\% of message types are relevant to state machine construction. Therefore, we manually select six representative symbols based on the protocol documentation to form the basic alphabet, improving the overall efficiency of the learning process.

\subsection{RQ2: Effectiveness of Mini-Batch Learning}
Table \ref{tab:combined_stats} demonstrates the efficiency of our approach compared to performing state machine learning over the full input symbol set. In experiments on protocols such as LightFTP, directly applying state machine learning to the full input symbol set fails to terminate within 12 hours; for other protocols, the learning process still requires several hours to complete. This high computational cost makes the naive approach impractical for semantic vulnerability detection when the alphabet size is large.
In contrast, both learning strategies based on extended alphabet sets are able to derive meaningful state machines within a bounded time, making them suitable for semantic vulnerability detection. Furthermore, the built-in basic state machine designed in our approach can further improve learning efficiency.

\subsubsection{Single-Symbol Mutation Priority Strategy}
As shown in Table \ref{tab:combined_stats}, the proposed single-symbol mutation–bounded strategy is able to complete state machine learning within hours—or even minutes in some cases. Compared to learning over the full input symbol set, the computational overhead is significantly reduced.
Although this approach may theoretically lose the ability to detect bugs that require the interaction of multiple message types, its low cost makes it well-suited for preliminary analysis of mutated symbols.

\subsubsection{Diverse-Symbol Mutation Priority}
As shown in Table \ref{tab:combined_stats}, our diverse-symbol mutation–prioritized strategy is able to learn dozens to hundreds of state machines within a 12-hour time budget. In contrast, for some protocol implementations such as LightFTP, state machine learning over the full input symbol set fails to complete within the same time limit. Our approach, however, produces a series of usable state machines under identical constraints, which can be directly leveraged for subsequent vulnerability detection.
Moreover, by controlling the window size, this strategy facilitates easier identification of the specific symbols responsible for anomalous behaviors when irregularities are observed in the learned state machines.

\subsubsection{Learning algorithm with the built-in basic state machine}
Table \ref{tab:combined_stats} also demonstrates the effectiveness of our built-in basic state machine strategy. For learning on the extended alphabet generated under the single-symbol mutation strategy, our method reduces the time cost by 32.5\% on average compared with the baseline method. For the diverse-symbol mutation scenario, our method learns 16.4\% additional state machines on average within 2 hours compared with the baseline.
\begin{table*}[htbp]
  \centering
  \caption{Learning Overhead and State Machine Count Statistics}
  \label{tab:combined_stats}
  \footnotesize
  \begin{tabular*}{\linewidth}{@{\extracolsep{\fill}} l c c c c c c c c}
    \toprule
    \textbf{Protocol} & \textbf{Implementation}
    & \multicolumn{3}{c}{\textbf{with set of extended alphabets}}
    & \multicolumn{4}{c}{\textbf{with built-in basic state machine}} \\
    \cmidrule(lr){3-5} \cmidrule(lr){6-9}
    & & \textbf{Full.} & \textbf{SP.} & \textbf{DP.}
    & \textbf{SP+.} & \textbf{Improvement.1 (\%)} & \textbf{DP+.} & \textbf{Improvement.2 (\%)} \\
    \midrule
    \multirow{3}{*}{FTP}
      & LightFTP    & $\times$  & 2225.5 & 636  & 1203.0   & 45.9  & 724 & 13.8  \\
      & ProFTPD     & $\times$  & 35250.1 & 18   & 21098.0 & 40.1  & 24  & 33.3  \\
      & Pure-FTPD   & $\times$  & 1601.2 & 815  & 1581.1  & 1.3   & 867  & 6.3  \\
    \midrule
    SMTP
      & Exim        & 22294.0   & 13733.0 & 35 & 4527.0  & 67.0  & 48 & 37.1     \\
    \midrule
    RTSP
      & Live555     & $\times$  & 7058.0 & 180   & 2241.1  & 68.2  & 198   & 10.0  \\
    \midrule
    \multirow{2}{*}{TLS1.3-Server}
      & wolfSSL     & 23478.2  & 497.4  & 527  & 387.0  & 22.2  & 585  & 11.0 \\
      & OpenSSL     & 13449.0  & 234.8  & 878  & 167.3 & 28.7  & 993  & 13.1 \\
    \midrule
    \multirow{2}{*}{TLS1.3-Client}
      & wolfSSL     & $\times$  & 884.5 & 757  & 768.1  & 13.1  & 823  & 8.7 \\
      & OpenSSL     & $\times$  & 754.8 & 857  & 708.5  & 6.1 & 974  & 13.6 \\
    \bottomrule
  \end{tabular*}

  \vspace{4pt}
  \scriptsize
  \parbox{\linewidth}{\centering
    \textit{Note:} Full.= Full input symbol set(Time required for learning);  $\times$=Learning non-terminated within 12 hours;\\
    SP. = Single-symbol Mutation Priority(Time required for learning); SP+. = SP combined with Learning algorithm with the built-in basic state machine; \\
    DP. = Diverse-Symbol Mutation Priority(Number of state machines); DP+. = DP combined with Learning algorithm with the built-in basic state machine; \\
    Improvement.1 = (SP - SP+ )/ SP $\times$ 100\%; 
    Improvement.2 = (DP+ - DP) / DP $\times$ 100\%;  \\
     membership query timeout: 1s; FIFO window size: 10 symbols; \\
    alphabet expansion iteration limit: 600s (10min); runtime budget: 12h (non-terminating if exceeded), 12h for diverse-symbol-priority setting.
  }
\end{table*}

\subsection{RQ3: Semantic bug detection}
\subsubsection{Authentication Bypass}
Our method successfully detects multiple authentication bypass vulnerabilities present in wolfSSL.

\begin{figure}[t]
  \centering
  \includegraphics[width=0.8\linewidth]{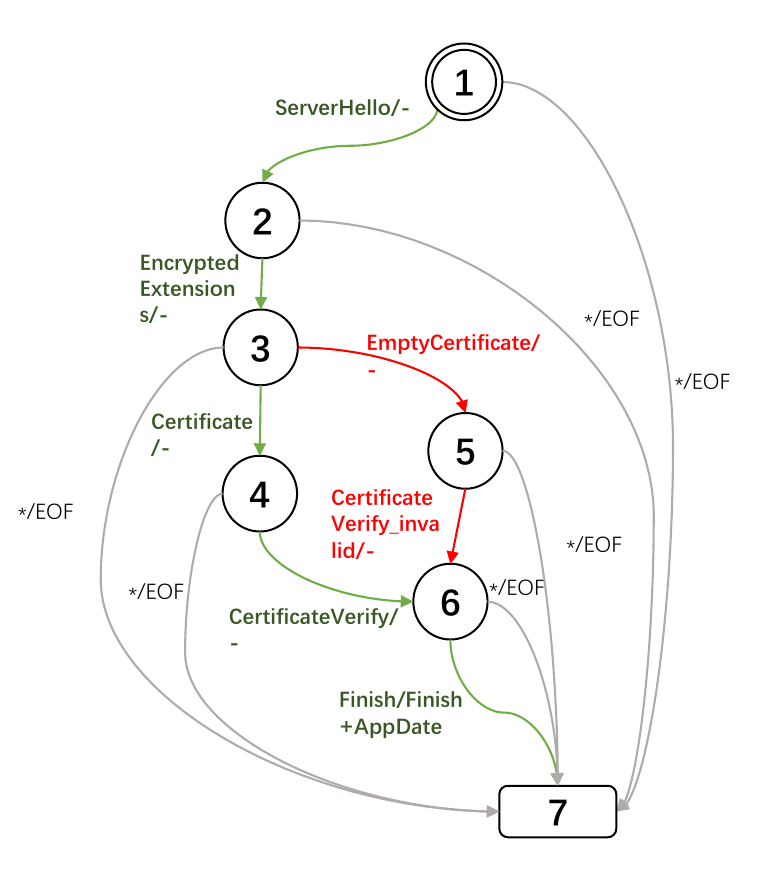}
  \caption{CVE-2021-3336, Vulnerable State Machine}
  \label{fig:cve2021}
\end{figure}

\textbf{CVE-2021-3336:}
Our method reproduces a critical authentication bypass vulnerability (CVE-2021-3336) in the wolfSSL client. As shown in Figure \ref{fig:cve2021}, an attacker can bypass server authentication and impersonate a server by sending an Empty Certificate message followed by a CertificateVerify message signed with an arbitrary RSA key. In our approach, this vulnerability is triggered by mutating a Certificate message using the Payload Nullification rule to generate an empty certificate, which is then added to the alphabet for state machine learning.

The successful reproduction of this vulnerability is primarily attributed to two factors. First, while formal certificate message specifications do not define an empty certificate as a valid type, our mutation operates at the structural level, resulting in a structurally valid empty certificate that is not recognized as an error by the client. Second, unlike manually injecting the empty certificate as a specific entry into the alphabet, our method generates this message through generic mutation. This represents a fundamental distinction from the approach used by Rasoamanana et al. to discover this vulnerability~\cite{Rasoamanana2022Systematic}.

\textbf{CVE-2022-25638:}
Our method also successfully reproduces the authentication bypass vulnerability CVE-2022-25638 presented in Section \ref{sec:Motivation Example}. As illustrated in Section 3, triggering this vulnerability requires the construction of two specifically crafted messages, namely EmptyCert and CV\_invalid. Both messages can be automatically generated through mutation under the proposed framework. Equipped with the designed Diverse-Symbol Mutation Priority strategy, our approach reliably detects the vulnerability within 12 hours across all five repeated experiments. From a probabilistic perspective, even adopting a purely random mutation configuration under identical experimental settings yields a detection probability exceeding 90\% within the same 12-hour duration.


%

\subsubsection{Specification Violation}
Our method has identified 10 unreported protocol non-compliance instances in LightFTP, ProFTPD, PureFTPD, and Live555. These semantic anomalies may lead to potential security risks, functional abnormalities, and information leakage in the protocol stack. Among the detected issues, three bugs have been fixed by the developers, and one has been assigned a CVE identifier.

\textbf{Duplicate and inappropriate return values:}
Our experiments reveal that certain commands in LightFTP and ProFTPD exhibit anomalous duplicate reply codes. Specifically, this issue occurs for the SITE command in LightFTP, as well as the STOR and LIST commands when the data connection has not been established. In ProFTPD, this issue affects the RNTO and SITE commands. 

Compared to traditional state machine learning approaches, we attribute these findings to two key features of our tool. First, mutated symbols enable the discovery of special command scenarios, such as issuing LIST and STOR without establishing a data connection. Second, even for symbols that do not require mutation, the FTP specification defines 34 commands, making state machine learning over the entire alphabet inherently time-consuming. In contrast, our mini-batch learning strategy, combined with a single-symbol–prioritized approach, allows for the rapid identification of symbols that exhibit semantic bugs.

This behavior constitutes a class of specification violations that can disrupt the normal operation of protocol implementations.
First, these duplicate reply codes exhibit implementation-specific characteristics across different FTP servers. This property can be exploited by attackers to quickly identify the underlying implementation, thereby enabling further targeted attacks—i.e., forming a form of state machine fingerprinting.
Moreover, this issue may lead to functional inconsistencies when interacting with different clients. Since clients are typically unaware of such bugs, they may misinterpret an extra reply code from a previous command as the response to a subsequent command, potentially rendering the entire connection unusable.
Notably, the incorrect handling of the RNTO command in Pure-FTPD has been assigned a CVE identifier.

\begin{table*}[htbp]
  \centering
  \caption{Statistics of Newly Discovered Semantic Bugs}
  \label{tab:protocol_lib_semantic_error_multi}
  \footnotesize
  \begin{tabular*}{\linewidth}{@{\extracolsep{\fill}} l p{5cm} p{5cm} c}
    \toprule
    \textbf{Implementation} & \textbf{Bug Description} & \textbf{Mutation Rule of Triggering Symbol} & \textbf{Status} \\
    \midrule
    LightFTP   & Duplicate and inappropriate return values & Payload Error Content Replacement (LIST) & Fixed \\
    LightFTP   & Duplicate and inappropriate return values & Payload Error Content Replacement (STOR) & Reported \\
    LightFTP   & Duplicate and inappropriate return values & $\times$ (SITE) & Reported \\
    LightFTP   & Non-compliant compatibility & Payload Error Content Replacement (TYPE) & Reported \\
    \addlinespace[2pt]
    ProFTPD    & Duplicate and inappropriate return values & $\times$ (RNTO) & Fixed with CVE Assigned \\
    ProFTPD    & Duplicate and inappropriate return values & $\times$ (SITE) & Reported \\
    ProFTPD    & Non-compliant compatibility & Header Error Content Replacement (USER) & Reported \\
    \addlinespace[2pt]
    Pure-FTPD  & Non-compliant empty response & Header Error Content Replacement (CWD) & Reported \\
    Pure-FTPD  & Non-compliant compatibility & Payload Error Content Replacement (PORT) & Fixed \\
    \addlinespace[2pt]
    Live555    & Non-compliant compatibility & Payload Error Content Replacement (PLAY) & Reported \\
    \bottomrule
  \end{tabular*}
\end{table*}
\textbf{Non-compliant compatibility:}
In our experiments, we discovered protocol specification violations in three libraries: LightFTP, ProFTPD, Pure-FTPD, and Live555. Specifically, all three implementations failed to strictly adhere to protocol specifications when processing certain erroneous messages. These issues can only be triggered by mutated symbols.


The specific bugs include: LightFTP handles the argument of the TYPE command loosely. It only validates the first character of the argument, ignoring any subsequent characters. For instance, the command "TYPE AW" is entirely invalid according to the protocol specification, yet LightFTP processes it identically to "TYPE A". ProFTPD treats the newline character following the command code of the USER command as a space, which does not comply with the FTP protocol standard. Pure-FTPD withholds all response codes when a quotation mark is appended to the end of the CWD command code, whereas the FTP specification mandates the return of a 500 error code for invalid commands in such cases. The Pure-FTPD server also handles the port specified in the PORT command in a permissive manner: when the given port exceeds the valid range, the server automatically applies a modulo operation to the port number. Live555, when processing the resource address in a PLAY command, disregards address validity and directly processes the content following a slash /. As a result, malformed commands such as "PLAY incorrect/ " are executed as valid instructions. 

Although these errors do not disrupt the execution of normal functionality, extending behavior to accommodate undefined or even non-conforming inputs without explicit documentation is still considered a semantic bug. Such deviations may lead to unexpected behavior for users of the protocol implementation.

\textbf{Non‑compliant empty response:}
We also observed undesired empty responses in Pure-FTPD during the processing of several commands. For the CWD command, when specific characters are injected into the command payload, the server returns a silent empty response. This issue also requires mutated symbols to be triggered. Although this behavior does not introduce functional failures, it may trigger unexpected runtime behaviors for upstream consumers of the protocol implementation.

\section{Limition and Future work}
Although automated mutation is adopted, hand-crafted mutation strategies exhibit inherent limitations: they fail to adequately cover bug-relevant input spaces and introduce redundant runtime overhead. Future work may integrate fuzzing techniques to generate more diversified and targeted mutation primitives.

Direct client-server interaction in our experiments incurs significant network I/O overhead and synchronous waiting, slowing the learning process. Future work should minimize this overhead or use multi-threading to mitigate synchronous waiting latency.
\section{RELATED WORK}

Testing stateful network protocols is a fundamental challenge in software testing and security analysis~\cite{Jian2024Fuzzing}. Existing research has mainly explored two complementary directions: protocol fuzzing and state machine learning. While both approaches have achieved substantial advances in exploring protocol implementations and discovering vulnerabilities, there are still some issues in efficiently identifying semantic bugs.

\textbf{Protocol Fuzzing:}
Existing approaches can generally be categorized into \emph{mutation-based} and \emph{grammar-based} fuzzing. Mutation-based protocol fuzzers extend traditional coverage-guided fuzzing to network protocols by generating test cases through input mutations and leveraging runtime feedback to guide exploration. Representative examples include AFLNet~\cite{pham2020aflnet} and its extensions such as NSFuzz~\cite{qin2023nsfuzz} and Nyx-Net~\cite{schumilo2022nyxnet}, which introduce various mechanisms to improve state exploration efficiency and fuzzing throughput. Another line of work explores \emph{grammar-based} or structure-aware fuzzing, which leverages input grammars to generate valid structured inputs (e.g., Superion~\cite{wang2019superion} and NAUTILUS~\cite{aschermann2019nautilus}).

Despite these advances, most protocol fuzzers primarily target memory-safety vulnerabilities such as buffer overflows and use-after-free errors, which are typically detected using runtime instrumentation tools like ASAN~\cite{serebryany2012addresssanitizer}. In contrast, detecting \emph{semantic bugs}, i.e., violations of protocol logic or state-machine behavior, often relies on manual inspection of abnormal interactions~\cite{Fiterau-Brostean2023AutomataBased}. Despite the existence of tools like DyFuzzing~\cite{Ammann2024DYFuzzing}, which aim to detect memory-corruption and semantic vulnerabilities in adversarial protocol environments, applying such tools to new protocol implementations requires substantial customization of the source program and testing harness, including protocol-specific instrumentation, which increases the deployment overhead for previously unseen protocols. Moreover, mutation-based fuzzing usually operates at the byte level, which may produce syntactically invalid messages that fail to exercise deeper protocol logic~\cite{Ammann2024DYFuzzing}. Consequently, fuzzing provides extensive input exploration capability but lacks mechanisms for systematically identifying semantic inconsistencies in complex protocol state machines. 

\textbf{State Machine Learning for Protocol Analysis:} 
Another research direction is \emph{active automata learning}, which aims to infer protocol state machines by systematically interacting with protocol implementations, and has been widely adopted in practical frameworks such as LearnLib~\cite{isberner2015learnlib}. Prior studies have successfully applied automata learning to analyze security protocols and reconstruct behavioral models of protocol implementations~\cite{deRuiter2015Protocol,tranvan2024mealy,Baumer2025FindingSSH}. These inferred models have proven instrumental for detecting semantic inconsistencies and vulnerabilities in widely deployed protocols such as TLS~\cite{Rasoamanana2022Systematic}.Some studies have improved state machine learning from different perspectives, such as improving the space efficiency~\cite{isberner2014ttt}, extending it to grey-box learning~\cite{StateInspector2022} and enhancing its ability to handle non-determinism~\cite{maehren2025towards}.

However, a fundamental bottleneck of many learning-based approaches lies in the construction of the input alphabet, which defines the set of protocol messages used during learning. In practice, alphabets are typically manually designed based on domain knowledge, restricting the explored input space and making large alphabets computationally expensive due to the large number of required queries. Recent work explores incremental alphabet construction for specific protocols such as TLS~\cite{maehren2025towards}, but these methods still rely on manually specified candidate messages. 
\section{Conclusion}
Ensuring the semantic correctness of stateful protocol implementations is crucial for system stability and security, yet existing detection approaches face limitations such as poor scalability and heavy reliance on expert effort. Addressing the core challenge of state machine learning in this domain—automated construction of input alphabets—this paper proposes a mutation-driven building paradigm. The framework integrates three key modules: LLM-assisted protocol configuration extraction, structured mutation, and mini-batch learning. Experimental results show that our approach effectively reproduces known semantic vulnerabilities, discovers previously unknown bugs (with three fixed and one assigned a CVE), and achieves efficient learning termination even for large candidate alphabets. This work provides a practical, extensible solution for semantic bug detection in stateful protocols, advancing the state of the art in protocol testing.
\section*{Data Availability Statement}
All source code supporting the findings of this paper are publicly and anonymously available
in a long-term archive with a DOI: \url{https://doi.org/10.5281/zenodo.19246632}.
\section*{AI Generative Tool Statement}
Generative AI tools were used to assist with language polishing and drafting of this paper.

\appendix
\section*{Appendix}

\makeatletter
\setlength{\@fptop}{0pt}        
\setlength{\@fpsep}{3pt}         
\setlength{\@fpbot}{0pt}         
\makeatother

\begin{algorithm}[H]
\centering
\caption{Single-Symbol Mutation Priority Algorithm}
\begin{algorithmic}[1]
\Require basic Alphabet $\Sigma_0$ ($\Sigma_0 \subseteq P$)
\Ensure set of Expanded alphabets $S=\{\Sigma_i\}$
\State \textbf{Initialization}: $S \leftarrow \emptyset$; $P_{\text{pending}} \leftarrow P-\Sigma_0$; $i=1$
\While{$P_{\text{pending}} \neq \emptyset$}
    \State Select an arbitrary letter $p \in P_{\text{pending}}$
    \State Compute all mutated forms of $p$: $\text{Mut}(p) = \{ f(p) \mid f \in F(p) \}$ and $\Sigma_i =\text{Mut}(p)\cup \Sigma_0$
    \State Expand the alphabet Set: $S \leftarrow S \cup \{\Sigma_i\}$
    \State Remove $p$ from pending set: $P_{\text{pending}} \leftarrow P_{\text{pending}} \setminus \{p\}$
    \State $i=i+1$
\EndWhile
\end{algorithmic}
\end{algorithm}

\begin{algorithm}[H]
\centering
\caption{Diverse-Symbol Mutation Priority Algorithm}
\label{alg:diverse_symbol_mutation}
\begin{algorithmic}[1]
\Require basic Alphabet $\Sigma_0$ ($\Sigma_0 \subseteq P$), Time threshold $T_{\text{th}}$, Max window size $W_{\text{max}}$
\Ensure set of Expanded alphabets $S$
\State \textbf{Initialization}: $\Sigma \leftarrow \Sigma_0$; $W \leftarrow \emptyset$; $\text{MutCount}(p) \leftarrow 0, \forall p \in \Sigma_0$; $P_{\text{pending}} \leftarrow P \setminus \Sigma_0$;
\Loop
    \State \textbf{Step 1:} Select symbol with least mutations
    \State $p^* \leftarrow \arg\min\limits_{p \in (\Sigma_0 \cup P_{\text{pending}})} \text{MutCount}(p)$;
    \State \textbf{Step 2:} Randomly generate one mutation
    \State Randomly select $f \in F(p^*)$; $q \leftarrow f(p^*)$; $\Sigma \leftarrow \Sigma \cup \{q\}$;
    \State $\text{MutCount}(p^*) \leftarrow \text{MutCount}(p^*) + 1$; $W \leftarrow W \cup \{q\}$;
    \State \textbf{Step 3:} Check thresholds
    \State $T_{\text{learn}} \leftarrow \text{TimeConsumption}(\Sigma)$;
    \If{$T_{\text{learn}} > T_{\text{th}} \lor |W| > W_{\text{max}}$}
        \State $w_{\text{old}} \leftarrow \text{head}(W)$; $W \leftarrow W \setminus \{w_{\text{old}}\}$; $\Sigma \leftarrow \Sigma \setminus \{w_{\text{old}}\}$;
    \EndIf
    \State \textbf{Step 4:} Update pending set
    \State $P_{\text{pending}} \leftarrow P_{\text{pending}} \setminus \{p^*\}$;
\EndLoop
\end{algorithmic}
\end{algorithm}

\begin{algorithm}[H]
\centering
\caption{State Machine Learning with built-in state machine}
\begin{algorithmic}[1]
\Require Basic Alphabet $\Sigma_0$, Expanded Alphabet $\Sigma$, Delta Alphabet $\Delta=\Sigma-\Sigma_0$
\Ensure Expanded State Machine $SM$
\State \textbf{Pre-Learning}: Perform $L^*$ algorithm on $\Sigma_0$, obtaining the base observation table $ot_0=(D_0,S_0,SA_0,F_0=\{(d,s)\to l \mid d\in D_0, s\in S_0 \cup SA_0 \})$
\State \textbf{Initialization}: Initialize new observation table $ot=(D,S,SA,F)$, assign $ot_0$ to $ot$
\State Expanding the Prefix Set: add $\Delta$ to $D$
\State Expanding the Precomputed Suffix Set: add $\Delta_{SA}=\{ s+\delta \mid \delta\in \Delta, s\in S\}$ to $SA$
\State Query the SUT and complete all missing observation table entries, add $\Delta_F=\{(d,s)\to l \mid (d,s)\in (D \times (S\cup SA))-(D_0 \times (S_0\cup SA_0)) \}$ to $F$
\State \textbf{Check the closedness of $ot$}
\State \textbf{Check the Consistency of $ot$}
\State \textbf{Constructing the hypothesized automaton}
\State \textbf{Query} the Minimally Adequate Teacher (MAT) for a counterexample
\If{Counterexample $ce$ exists}
    \State \textbf{Add} $ce$ to the observation table
    \State \textbf{Goto} \textbf{Line 6} \Comment{Return to check}
\Else
     \State  Convert $ot$ to $SM$
    \State \textbf{Return} $SM$ \Comment{Output automaton}
\EndIf
\end{algorithmic}
\end{algorithm}
\bibliographystyle{ACM-Reference-Format}
\bibliography{bibfile}
\end{document}